\documentstyle[12pt]{article}
\begin{document}
\date{}
\large

\title{ Cherenkov radiation of magnon and phonon by
the slow magnetic monopole.}
\author{I.V.Kolokolov, P.V.Vorob'ev}
\date{Budker Institute of Nuclear Physics (BINP)\\
630090 Novosibirsk, Russia, vorobyov@inp.nsk.su, kolokolov@inp.nsk.su}

\maketitle

\begin{quotation}
 {\it The Cherenkov radiation of  magnons  at  passage  of the heavy  slow
magnetic  monopole through an ordered magnetic matter
is considered. Also the Cherenkov radiation  of  phonons  at  monopole
movement in  medium is discussed}.

\end{quotation}

\section { Introduction }
The concept of magnetic monopole has been entered into modern
physics in
1931 by Paul Dirac \cite {Dir}. He supposed the existence of
isolated magnetic charge $g$ ---  $ge=\frac{n}{2} \hbar c$,
where $e$  ---  electrical  charge,  $\hbar$  --- Planck constant,
$c$ --- the light speed , $n = \pm 1,2...$ --- integer.
Numerous  and  unsuccessful  attempts  of  experimental    search
for magnetic monopole on accelerators \cite{Barkov,FNAL}
and in cosmic rays \cite{Groom,Klapdor} were done since  then.

The new interest to this problem has arised in 1974, when Polyakov
\cite {Pol} and 't~Hooft \cite {tHooft} have shown, that such objects
exist as solutions in a wide class of models with spontaneously broken
symmetry.

The registration of Dirac's  monopole  or  evaluation  of  their
flux limit  will  the essential  contribution  to
construction of Grand Unified Theory, as well as  give a pulse to the
decision of many problems in astrophysics.
Therefore the study of various  mechanisms  of  the Dirac's  monopole
interaction with  medium is important as  from   fundamental, as  from
applied point of view.
We consider here the Cherenkov radiation of magnons at  passage  of
heavy slow moving magnetic monopole through an  ordered  magnetic
matter and discuss also the Cherenkov radiation  of  phonons  at  monopole
movement in  matter.

\section {Excitation of spin wave Cherenkov radiation by the heavy magnetic
monopole.}

As well known, the slowly moving heavy monopole cannot emit
usual Cherenkov radiation in ferromagnetic media,  because  the  phase
speed of electromagnetic waves is of the order $c/10$  and
always much faster than the monopole speed.

We  shall  consider  the slow monopole passage through an ordered
magnetic matter. In such case a  main  mechanism  of  kinetic
energy loss is the Cherenkov radiation of magnons. This is because the
magnon phase velocity  reaches zero  and  the  coupling of
monopole to magnons is linear and large \cite {VKI}.

For definiteness we shall consider a ferromagnet, but the evaluations
below are of more general character.

Magnon's Hamiltonian  in presence of magnetic field of a moving monopole
can be written in the form

\begin{equation}
H = \sum_ { \bf k } \hbar \omega_{ \bf k} a^{\dagger}_{\bf k} a_{
 \bf k } + \sum_ {\bf k} \left (f_{\bf k} e^{ -i\Omega_{ \bf k} t}
a^ {\dagger}_{ \bf k} + c.c \right )~,
\label {sw1}
\end{equation}
where $ a^ { \dagger}_{\bf k}$ --- operator of  a  magnon  birth
with a wave vector ${\bf k}$, $\omega_ { \bf k}$---his  dispersion
law, $\Omega_ { \bf k } = {\bf  kv }$, ${\bf v}$---vector  of  a
monopole speed and $f_{\bf k}$  ---  coupling  factor  of  a  monopole
magnetic field $ {\bf B}=g {\bf \nabla} \frac {1} {r}$  with
magnon.

The magnon energy, radiated in a unit of time, is
\begin{equation}
\epsilon =\frac {2\pi}{\hbar}\sum_{\bf k} \omega_{\bf k} {\left| f_{ \bf k}
\right |}^2 \delta (\Omega_{\bf k}-\omega_{ \bf k} )~.
\label {sw2}
\end{equation}
Let the monopole speed $\bf{v}$ be directed along the direction  of  the
spontaneous magnetization, along Z-axis. $Z$ \footnote
{ the general case is investigated absolutely  similarly  and
the result differs only by a factor close to 1.}.
Then
\begin{equation}
F_{\bf k}=\frac {4\pi g\mu_B} {a^{3 /
2} \sqrt {V}} \sqrt {\frac {S} {2}} \frac {k_x-ik_y } {k^2}~,
\label {sw3}
\end{equation}

here $a$ is the lattice constant, $V$ is the sample volume, $S$ is the
spin size on the node and  $\mu_B$ is the Bohr magneton.

Taking into consideration (\ref{sw3}) the equation for
$\epsilon$ can be written as

\begin{equation}
\epsilon = \frac { 2 g^2 \mu_B^2 S} {a^3 \hbar} \int d^3 {\bf k} \omega_
{\bf k} \frac {k_x^2 + k_y^2} {k^4} \delta (k_z v-\omega_{\bf k} )~.
\label {sw4}
\end{equation}
The integration in (\ref {sw4}) is performed on the first Brillouin zone.

If $v \ge u$, where $u$ --- magnon speed near a border of Brillouin zone,
then the magnons with large $\bf k$ are essential. Then

\begin{equation}
\epsilon \simeq \frac {\bar {\omega} g^2 \omega_M} {v}~,
\label {sw5}
\end{equation}
where the frequency $\omega_M = \frac { 4\pi\mu_B^2 S } { \hbar a^3 } $
characterizes magnetization of media \cite {Gur},

\begin{equation}
\bar { \omega } =\frac {1} {2\pi} \int \frac {d^2\bf {k_{\bot}}} {k_
{\bot}^2 }\omega_{k_{\bot}}~,
\label {sw6}
\end{equation}

here $\bf { k_{\bot}} = (k_x, k_y)$, and $\bar {\omega}$ has the value about
maximal frequency of magnons.

For $g^2 \simeq 4700 \cdot e^2$ we obtain

\begin{equation}
\epsilon \simeq 10^3 \cdot Ry \cdot \omega_M (\bar {\omega} \tau)~,
\label {sw7}
\end{equation}
where $\tau=a/v$ is the characteristic time of interaction.

The typical values for magneto-ordered dielectrics are such:
$\bar {\omega } \simeq 10^{-13}s^{-1 }$, $\omega_M \simeq 10^{-11}
s^{-1}$ and for $v/c \simeq 10^{-4}$ --- $\epsilon \simeq 10^{14}
$eV/s, that corresponds to losses per unit of length:
$$\frac {dE} {dl} \simeq 10^{8}~eV/cm$$

From (\ref{sw5}) it is clear, that the losses $\epsilon$ and $dE/dl$
grow with slowing down of monopole. When the speed $v$ becomes
$v<u$, the main contribution to losses contribute the magnons
"from the bottom" of the spectrum.
For them, $\omega_{\bf k}=\omega_{ex} (ak)^2$, where $\omega_{ex}$ is the
frequency, characterizing the exchange interaction \cite {Gur,Ahi}~, and the
expressions for losses acquire the shape:
\begin{equation}
\epsilon = g^2 \frac {\omega_M v} {4 \omega_{ex}a^2}~;
\label {sw8}
\end{equation}

\begin{equation}
\frac {dE} {dl} = \frac {\epsilon} {v} = g^2\frac {\omega_M}
 {4\omega_{ex}a^2}~.
\label {sw9}
\end{equation}

As one can see, the energy losses per unit of a length approach a constant
with reduction of the monopole speed. The characteristic values will be
$\omega_M / \omega_{ex} \simeq 10^{-2}$,
and for $v/c \simeq 10^{-4}$ have $$a \simeq 10^{-8}~cm~;$$
$$\frac {dE} {dl} \simeq 10^{8}~eV/cm~.$$

We'd like to specially note, that the square-law of magnon  dispersion
leads to non-trivial spatial structure of Cherenkov radiation field of
spin waves. As it is, usually, the structure of a radiation  field  is
similar to a shock wave and advancing the charge  radiation  is  away.
For the square-law dispersion of the  radiation field  advances  charge
and is not equally to zero before charge. It is due to  that  for  the
square-law dispersion the group velocity of a wave is more then  phase
(and more than velocity of a charge movement).

From these evaluations it is clear, that a level of energy losses of a slow
magnetic monopole in magneto-ordered matter could be compared to
ionization losses of a fast monopole. This opens new opportunities for
construction of detectors of monopoles in the range of $v/c < 10^{-4}$.
The conversion of spin waves to electromagnetic \cite {Ahi} permits to
detect a monopole passing through a magnetic layer by traditional means.

\section { Excitation of Cherenkov acoustic (phonon) radiation
by heavy magnetic monopole.}

For valuation of energy losses by radiation of sound waves (excitation
of phonons) by the monopole moving in the isotropic matter,  we  shall
write the Hamiltonian of an elastic system  in  an  external  field  as
follows
\begin{equation}
H=\sum_{n} \frac {{\bf p}_{n}^2} {2M} +
\frac {a} {2} \sum_{n,\Delta} ({\bf x}_{n}-{\bf x}_{n + \Delta} )^2 +
\sum_{n} {\bf F}_{n} (t) {\bf x}_{n}~.
\label {aw1}
\end{equation}
Here $n + \Delta$ are the numbers the closest neighbors to the node $n$,
\begin{equation}
{\bf F}_{n} (t) = {\bf F} ({\bf r}_{n} - {\bf v} t)~.
\label {aw2}
\end{equation}

We shall estimate the strength of the force ${ \bf F} ({ \bf r}_{n})$,
acting from the monopole to the given  node,  as  follows.  First,  we
shall assume that this force is located on one node (it's  short-range
nature allows this):
\begin{equation}
{ \bf F (r_{n})} = {\bf F} \delta_{n0}~.
\label {aw3}
\end{equation}

Secondly, at rest this force causes deformation, and the affected node
is shifted by
\begin{equation}
\delta a \sim \frac {F} {A}~,
\label {aw4}
\end{equation}
and, assuming the deformation energy to be $\epsilon_{def} \sim
A\delta a^2$, we have the force as
\begin{equation}
{\bf F} \sim A\sqrt { \frac{\epsilon_{def}} {A}} \sim
\sqrt{\epsilon_{def} A}~.
\label {aw5}
\end{equation}

The Hamiltonian in (\ref{aw1}) can be  expressed  completely  similar  to
(\ref{sw1}):

\begin{equation}
H = \sum_{ \bf k} \hbar \omega_{ \bf k} a^{\dagger}_{\bf k}a_{\bf k} +
\sum_{\bf k} \left (f_{\bf k} e^{-i\Omega_{ \bf k} t }
a^{ \dagger }_{ \bf k} + c.c \right ) ~,
\label{aw6}
\end{equation}
but now:
$ a^{\dagger}_{ \bf k}$ is the operator of phonon creation with the
wave vector
${ \bf k}$, $\omega_{\bf k} $ is it's dispersion,
$\Omega_{ \bf k } = {\bf kv}$, ${ \bf v}$ is the vector of monopole speed,
and $f_ { \bf k }$ is the coupling factor of the magnetic monopole field
${\bf B } =g\nabla \frac {1} {r} $ with the phonon. We shall write the
following expressions for them:

\begin{equation}
F_{\bf k} =
\frac {1} {2i} \hbar ^{1/2} \frac {1} {\sqrt {N}}
\frac {F} {( D_{\bf k} M)^{1/4}}~,
\label {aw7}
\end{equation}

\begin{equation}
\omega_ { \bf k } = \sqrt { \frac { 2 D_ { \bf k }} { M }}
\label{aw8}
\end{equation}

\begin{equation}
D_{\bf k} = \frac {A} {2} \sum_{\bf {\Delta}} \left | 1-e^{i {\bf k \Delta
}} \right | ~.
\label{aw9}
\end{equation}

Accordingly, the energy of phonons, radiated in a unit of time, is equal to
\begin{equation}
\epsilon = \frac {a^3} {4 (2\pi)^2} \int {d^3 {\bf k} \frac {F^2}
{(D_{k} M)^{1/2}}\omega_{\bf k} \delta (k_z v-\omega_ {\bf k})}~,
\label{aw10}
\end{equation}
where $a$ is the constant of the lattice $a^3=V/N$, $V$ is the
sample volume.

For the essentially supersound monopoles, integration over $dk_z$
gives the factor $1/v$, and the integral (\ref{aw10}) results to:
\begin{equation}
\epsilon = \frac {a^3} {4 (2\pi)^2 } \frac {1} {v} \int {d^2{\bf k}_{\bot}
\frac { F^2} {(D_{{ \bf k}_{\bot}} M)^{1/2}}
\omega_{{\bf k}_{\bot}}} =
\frac {a} { 2\sqrt {2} v} \frac {F^2} { M }~.
\label {aw11}
\end{equation}

Using the evaluation from Eq.(\ref{aw5}), for $F$ we shall obtain:
\begin{equation}
\epsilon \simeq \epsilon_{def} \frac {a} {v} \frac {A}{M}~.
\label {aw12}
\end{equation}

Now from decomposition (\ref{aw9}) at small ${\bf k} $ and using
Eq. (\ref{aw8}) it is possible to express $A/M$ through speed
of sound. As a result (\ref{aw12} ) acquires the shape
\begin{equation}
\epsilon \simeq \epsilon_{def} \frac {1} {Z} \frac { c_{s}} {v}
\frac {c_{s}} {a} \simeq \epsilon_{def} \frac { c_{s}} {v} {
\bar {\omega}_a}~,
\label {aw13}
\end{equation}
where ${\bar {\omega}_a}$ is the cut-off frequency for phonons,
$Z$ is the number of nearest neighbors.

If $\epsilon_{def} \sim Ry$, $c_{s}/v \sim 0.1$ and
${\bar {\omega}_a} \sim 10^{13}~s^{-1}$
(${\bar {\omega}_a}$ is about the Debye temperature in energy units),
then:
$$\epsilon \simeq 10^{13} eV/s,$$
$$\frac {dE} {dl} \simeq 10^{7} eV/cm,$$
which is a little less than loss by radiation of magnons.\\

We would like to thank L.M.~Barkov, V.V.~Ianovski and I.B.~Khriplovich
for a number of useful discussions.

\end{document}